\documentclass[lettersize,journal]{IEEEtran}
\usepackage{amsmath,amsfonts}
\usepackage{algorithmic}
\usepackage{algorithm}
\usepackage{array}
\usepackage[caption=false,font=normalsize,labelfont=sf,textfont=sf]{subfig}
\usepackage{textcomp}
\usepackage{stfloats}
\usepackage{url}
\usepackage{verbatim}
\usepackage{graphicx}
\usepackage{cite}
\usepackage{xcolor,cite,etoolbox}
\begin{document}

\title{A Planar Super-Directive Huygens Antenna with Enhanced Aperture Efficiency}

\author{Muhammad Rizwan Akram,~\IEEEmembership{Member, IEEE,} and Abbas Semnani,~\IEEEmembership{Senior Member, IEEE}
\thanks{The authors are with the Department of Electrical Engineering and Computer Science, The University of Toledo, Toledo, Ohio 43606, USA. (email: muhammadrizwan.akram@utoledo.edu; abbas.semnani@utoledo.edu). This work was supported by the Office of Naval Research (ONR) under Grant N00014-21-1-2449.}
}

\markboth{IEEE Antennas and Wireless Propagation Letters,~Vol.~x, No.~xx, 2025}%
{Shell \MakeLowercase{\textit{et al.}}: A Sample Article Using IEEEtran.cls for IEEE Journals}


\maketitle

\begin{abstract}
The magnetic dipole in antenna design is conventionally realized using an electrically small loop, which can be paired with an electric dipole to form an electrically small Huygens’ antenna. However, such configurations typically exhibit low radiation efficiency and a theoretical directivity limit of 4.8 dBi. This paper introduces a magnetic dipole formed by a $0.5\lambda_r$ slot positioned adjacent to a printed dipole twice its length. The resulting combination of electric and magnetic dipoles produces a highly directive radiation pattern with a gain exceeding that of a uniformly illuminated aperture of comparable size. A prototype operating at 4.5 GHz achieves a measured directivity of 8.37 dBi, in close agreement with analytical and numerical predictions. The proposed high-gain, low-profile antenna features PCB compatibility, a simple feeding structure, and low side- and back-lobe levels—making it a promising candidate for compact, energy-efficient, and secure wireless systems.  
\end{abstract}


\section{Introduction}
\IEEEPARstart{T}he pursuit of superdirective antennas has remained a major challenge for nearly a century. Harrington investigated the fundamental relationship between antenna size and gain using spherical wave functions \cite{harrington1958gain}. The theoretical maximum gain for an antenna enclosed within a sphere of radius $r$, known as the Harrington limit, is expressed as
\begin{equation}
    D_{max}^{H} = {[N]}^2+2N.
    \label{equ1}
\end{equation}
Here, $N = kr$ is considered to relate to the physical size of the antenna, indicating the approximate transition of rapid and slowly varying waves of the Hankel function, with $k$ being the wavenumber and $r$ the radius of the sphere in which the antenna can be enclosed. For large $kr$, the gain limit simplifies to ${(kr)}^2$, which corresponds to the gain of a uniformly illuminated circular aperture. Although Harrington's limit is often associated with superdirectivity, a widely accepted definition by Hansen \cite{hansen1981fundamental} states that a superdirective antenna exhibits greater directivity than a similar-sized antenna with uniform excitation. The directivity of a uniformly illuminated aperture is given by
\begin{equation}
    D = \frac{4\pi A}{{\lambda_0}^2},
    \label{equ2}
\end{equation}
where $\lambda_0$ is the free-space wavelength, and \textit{A} represents the physical size of the aperture.

Oseen et al. \cite{oseen1922einsteinsche} first considered the theoretical possibility of superdirectivity from arbitrarily small sources, followed by Riblet \cite{riblet1948note}. Uzkov \cite{uzkov1946approach} demonstrated that a linear array of $N$ isotropic radiators could achieve a maximum directivity of $N^2$. The $N^2$ is an extraordinary directivity that can be achieved compared to the directivity of "N", realized by isotropic radiators half a wavelength apart. Several attempts have been made to demonstrate a superdirective antenna array. Bloch et al. experimentally demonstrated four half-wavelength dipoles spaced 0.2$\lambda$ apart to form a superdirective array \cite{bloch1953}. Altshuler et al. \cite{altshuler2005monopole} demonstrated an increase in gain of 6 dB compared to a single element of two monopoles 0.2$\lambda$ apart. Other subsequent works include an electrically small end-fire array by O'Donnell et al. \cite{o2006electrically} and Yaghjian et al. \cite{yaghjian2008electrically}, a two-element superdirective antenna by Best et al. \cite{best2008impedance}, and the 4-element superdirective array by Clemente et al. \cite{clemente2015design}. These demonstrations utilize end-fire configurations to place antenna elements in the array. This approach is susceptible to narrow tolerances and requires a delicate balance of phase and amplitude to achieve a superdirective radiation pattern. 

The size restrictions stipulated by Harrington in (\ref{equ1}) do not apply to electrically small antennas (ESAs), and it is theoretically possible to achieve infinite gain, depending on the excitation of several modes. However, the Harrington limit cannot be applied for ESAs because the assumption $N\ge kr$ is not valid for ESAs. Gayi \cite{geyi2003physical} studied the Q of the antennas and defined the maximum gain limit for the ESAs. Kwon \cite{kwon2005radiation} obtained the expression for electrically small crossed electric and magnetic dipoles and showed that their gain could be as high as 3 (4.8 dB). Yaghjian \cite{yaghjian2009increasing} discussed the maximum achievable gain for the Huygens' array and the possible realization of Huygens' radiators. Pigeon et al. \cite{pigeon2014miniature} investigated the feasibility of maximizing the directivity of miniaturized directive antennas using available techniques.

The development of Huygens sources for antenna applications remains challenging, primarily because of the difficulty of simultaneously achieving impedance matching for both the electric and magnetic dipole components. Several practical implementations have been proposed in the literature. For example, Ziolkowski et al. \cite{jin2010metamaterial,ziolkowski2015low,ziolkowski2019custom,tang2016design,lin2017electrically} employed combinations of loop and wire structures, while Niemi et al. \cite{niemi2012electrically} utilized dual omega particles to realize Huygens-type radiation. Recently, Ziolkowski \cite{ziolkowski2023superdirective} numerically investigated the feasibility of a mixed multipole antenna at 30 GHz, utilizing a layered structure that achieves high gain. Later, a practical design was demonstrated as an electrically small Huygens' quadruple antenna, exploiting two pairs of Huygens' sources \cite{zhang2024superdirective}.

This paper presents a novel planar Huygens antenna topology that enables super-gain radiation, offering the potential for enhanced SNR in the desired direction even under low-power operation. The antenna is low-profile, compatible with standard PCB fabrication processes, and achieves a realized gain of approximately 1.2$\times$ that of a uniformly illuminated aperture of comparable size. This performance is achieved through the constructive in-phase excitation of a magnetic dipole and a full-wavelength electric dipole, pushing the gain performance toward the Harrington directivity limit. Contrary to recent superdirective antennas such as the end-fire antenna array \cite{lynch2024super} and the hybrid end-fire antenna (Huygens quadruple) \cite{zhang2024superdirective}, the proposed Huygens' antenna does not require any complex feeding, can operate with nearly 100$\%$ radiation efficiency, and can be easily scaled to higher gains using low permittivity dielectrics.

\section{Antenna's Design}
\begin{figure}[b!]
\centering
\includegraphics[width=0.6\linewidth]{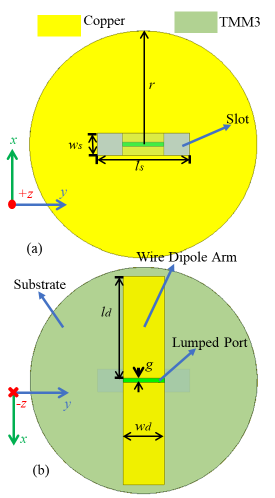}
\caption{Topology of the proposed Huygens' superdirective antenna (a) $y$-aligned slot (magnetic dipole) on the top side (b) $x$-aligned printed electric dipole on the bottom side.}
\label{Fig1: topology}
\end{figure}

The geometry of the proposed Huygens' antenna is shown in Fig.~\ref{Fig1: topology}. The antenna consists of a cylindrical disk made of Rogers TMM3 board with dielectric constant $\epsilon_r = 3.27$ and loss tangent $tan\delta = 0.002$. The disk thickness is $h = 2.54$ mm, with 0.5 oz copper layers on both sides. A slot of length $\l_s = 20.2$ mm and width of $w_s = 5$ mm is etched from the top side. On the bottom side, two arms of the dipole are printed along the y-axis, separated by a gap $g = 1$ mm for feeding. The length and width of each arm are $l_d = 22.5$ mm and $w_d = 9$ mm, respectively. The lumped port is used to numerically feed the dipole arm and is then replaced by an SMA connector in the fabricated prototype. To match impedance with magnetic and electric dipoles simultaneously, each arm of the printed dipole is nearly the same size as the slot, resulting in a printed dipole that is twice the size of the slot, with a total effective electrical length of 1$\lambda_r$. This leads to the formation of a full-wavelength electric dipole mode in the printed dipole and a magnetic dipole mode in the cross slot. The two dipoles radiate in phase, resulting in very high directivity along the +z-axis and cancellation of radiation in the backward direction, achieving a high front-to-back ratio of nearly 12 dB.

\begin{figure}[!]
\centering
\includegraphics[width=0.95\linewidth]{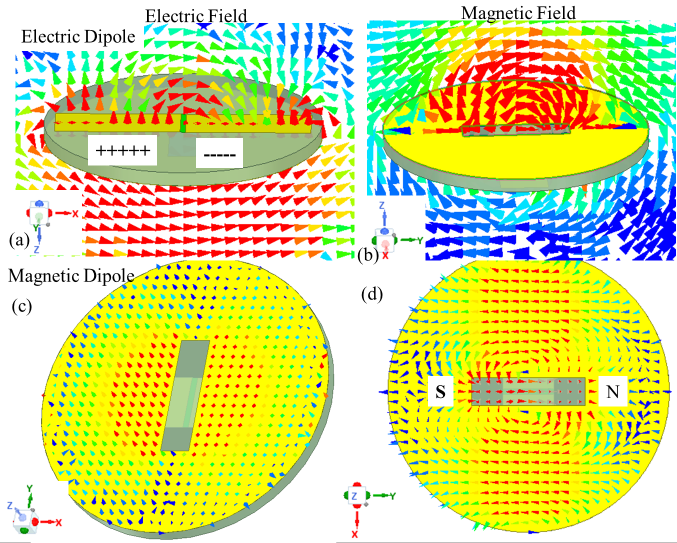}
\caption{The electric fields (a,c) and magnetic fields (b,d) over the designed Huygens' antenna, forming $x$-oriented electric dipole (a) and $y$-oriented magnetic dipole (d).}
\label{Fig3: fields}
\end{figure}

 The electric and magnetic fields of the proposed antenna topology, as shown in Figs. \ref{Fig3: fields}(a) and (b) for the electric dipole and in Figs. \ref{Fig3: fields}(c) and (d) for the magnetic dipole. The electric field over the two arms of the printed dipole indicates the accumulation of positive charges on one arm and negative charges on the other, forming a typical pattern of an electric dipole around which the magnetic field is revolving. The higher strength of the electric and magnetic fields along +$z$ indicates the constructive interference of both dipoles to form Huygens' source. The slot acts as a magnetic dipole, as confirmed by the formation of south and north poles along the $y$-axis and the rotation of the electric field around the slot, which is the typical behavior of a magnetic dipole. The $y$-directed magnetic dipole and the $x$-directed electric dipole form Huygens' antenna.

\section{Results Discussion}
\begin{figure}[!]
\centering
\includegraphics[width=0.8\linewidth]{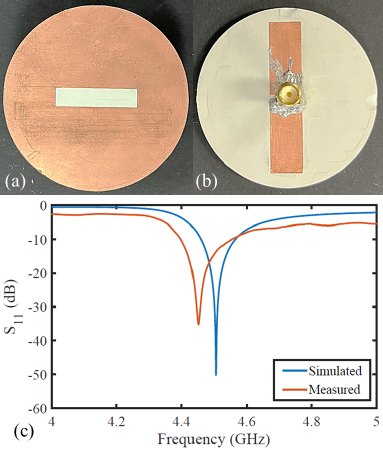}
\caption{Fabricated Huygens' antenna: (a) top side, (b) bottom side, and (c) its reflection response.}
\label{Fig6: Tun}
\end{figure}

The proposed Huygens' antenna is numerically verified using high-frequency simulation software (HFSS) and then fabricated from Rogers TMM3 boards using a PCB prototyping machine. The SMA connector feeds the EM energy through a vector network analyzer (VNA) with the center pin of the SMA connected to one arm and the ground connected to the other arm. The fabricated and assembled boards are shown in Fig.~\ref{Fig6: Tun} together with their simulated and measured reflection response. The antenna's operating frequency is 4.472 GHz in the experiments and 4.505 GHz in the measurements. A deviation of 33 MHz is observed compared to simulations, mainly attributed to the slight additional length introduced by the center pin of the connector. The dips, particularly on the left side of the central frequency, are mainly caused by a balanced-to-unbalanced system, which can typically be resolved using a balun. The reflection coefficient below -50 dB indicates that the antenna is very well matched at the operating frequency with almost 100\% energy coupled to the antenna.

\begin{figure}[!]
\centering
\includegraphics[width=0.8\linewidth]{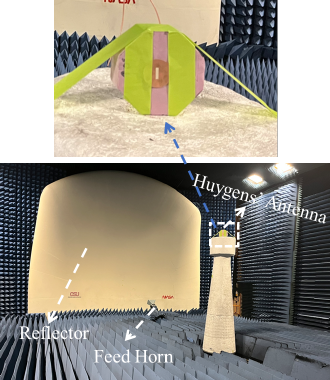}
\caption{Antenna's far-field measurement setup.}
\label{Fig6b: setup}
\end{figure}

\begin{figure}[!]
\centering
\includegraphics[width=0.95\linewidth]{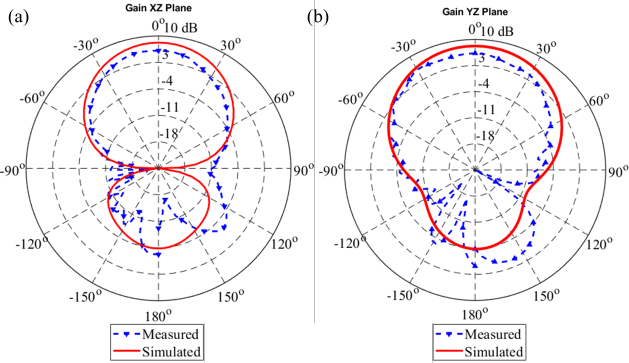}
\caption{2D radiation pattern along $XZ$-plane and $YZ$-plane.}
\label{Fig7: 2dplots}
\end{figure}

Since a low dielectric of TMM3 is used for high radiation efficiency, a simulated radiation efficiency of 97\% is obtained with a directivity of 8.37 dBi. Since there is no loss in impedance matching and minimal ohmic loss, the realized gain is close to the maximally achievable directivity at 8.27 dBi, only 0.1 dB lower than the maximum directivity. The antenna is experimentally characterized in a compact range anechoic chamber. The experimental setup is shown in Fig.~\ref{Fig6b: setup}. The antenna under test is placed on a rotating column, which is then illuminated by the plane wave from the reflector. A standard horn antenna feeds the reflector. The simulated and measured 2D radiation pattern in the XZ and YZ planes is shown in Fig. \ref{Fig7: 2dplots}. The experimental and numerical results are in good agreement. The peak gain obtained is nearly 1.6 dB less than the measured value, primarily due to the slight variation in the perfect excitation of electric and magnetic modes resulting from the additional length introduced by the connector pin. Minor inaccuracies can also be caused by the cable on the back of the antenna, affecting the backward cancellation of the fields. Furthermore, the reference gain antenna has a tolerance of $\pm$0.5 dB. 

\begin{figure}[!]
\centering
\includegraphics[width=0.95\linewidth]{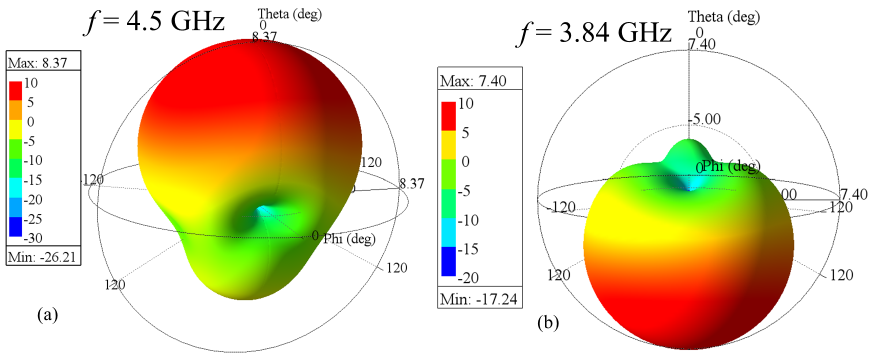}
\caption{3D simulated directivity, (a) forward propagation mode, and (b) backward propagation mode of Huygens' antenna at different frequencies.}
\label{Fig8: directivity}
\end{figure}

The 3D plot for simulated directivity is shown in Fig.~\ref{Fig8: directivity}. The peak directivity reaches 8.37 dBi, accompanied by a front-to-back ratio of approximately 12 dB. The high front-to-back ratio indicates nearly equal matching with electric and magnetic dipoles. The overall antenna size of 1$\lambda_r$ increases the gain to a much higher value than could be achieved by the Huygens' ESA. Due to the symmetrical nature of the fields in the front and back directions for Huygens' source, the broadside direction can be easily switched by reversing the polarity of either the electric or magnetic fields. In the current design, backward radiation appears at 3.86 GHz, slightly below the main operating frequency of 4.45 GHz. The maximum realized directivity of 8.37 dBi is 120\% of that of the uniformly illuminated antenna of the same size calculated using (\ref{equ2}). The maximum directivity with such a size can be 10.11 dBi, according to Harrington (\ref{equ1}). The realized directivity is only 1.74 dB, less than the maximum Harrington limit.

\section{Design Flexibility}
\begin{figure}[!]
\centering
\includegraphics[width=0.95\linewidth]{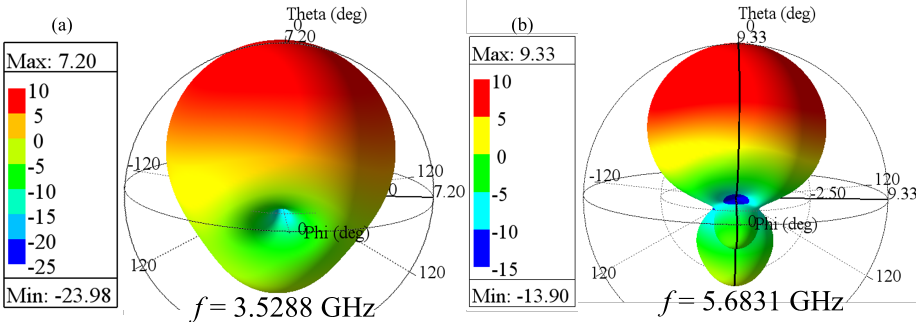}
\caption{3D simulated directivity for (a) TMM6 and (b) RT/Duroid 5880 boards as the  antenna's substrate.}
\label{Fig9: direc}
\end{figure}

The proposed approach can be easily scaled to any frequency as long as the 0.5 ratio is maintained between the length of the slot and the total dipole length. Moreover, lower dielectric constants can enhance antenna gain as a result of increased electrical size. In this regard, three different dielectrics, that is, TMM3 with $\epsilon_r  = 3.27$, TMM6 with $\epsilon_r  = 6$, and RT/Duroid 5880 with $\epsilon_r  = 2.2$ are used as substrate to investigate the directivity in the context of the Harrington limit, keeping the overall physical size of the antenna the same, that is, it can fit in a sphere of radius 25 mm. The results for TMM3 are already presented in the previous section. The directivities are plotted in Fig.~\ref{Fig9: direc} for the other two. The antenna can achieve 9.33 dBi of directivity when RT/Duroid is used as a substrate because of the increased electrical size resulting from the frequency shift upward caused by the low permittivity of the dielectric. The maximum directivity corresponding to $ka$ = 2.98, according to Harrington (\ref{equ1}), is 11.7 dBi, resulting in a 2.37 dB gap between the realized and theoretical limits.

\begin{table*}[!]
\centering
           \caption{Comparison with the State-of-the-Art Superdirective Antennas}
           \small
            \renewcommand{\arraystretch}{1.2}
            \begin{center}
                \begin{tabular}{c c c c c c c c c c}            \hline        \hline
& Topology  & $f_c$ & $ka$ & $D_{max}^H$$^a$ & D  & $\eta_{tot}$  & $G_R$ & Complexity$^b$ \\
&   &  (GHz) &  &   (dBi) & (dBi) &  ($\%$) &  (dBi) &   \\ \hline

                   \cite{altshuler2005monopole} & 2-element monopole array & 0.4 & 7.11  & 18.1  & 9.5 &  -& - & High\\

                   \cite{clemente2015design} & 4-element dipole array & 0.868 & 1.7584  & 6.61  & 11.7 & - & -18 & High\\

                  \cite{kim2012superdirective} & 2-element magnetic dipole array & 0.435 & 3.28  & 12.35  & 9.9 & 55 & - & High\\

                  \cite{debard2022maximum} & 3- and 4-element dipole array & 0.85, 0.85 & 1.46, 2.06 & 5.077, 8.36  & 9, 10.3 & - & -1.6, 2.1 & High\\

                  \cite{haskou2015design} & 2-element printed half loop array  & 0.901 & 1.22  & 6  & 6.8 & 40 & - & High\\

                  \cite{lynch2024super}  & 2-element strip dipole array & 3.5 & 1.55 & 7.44   & 6.5  &  99.3  & 6.3 & moderate\\
                  \cite{ziolkowski2023superdirective}  & UMMA (Simulation only) & 28.3 & 2.11 & 9.38   & 9.52  &  77.9  & - & moderate\\

                  \cite{zhang2024superdirective}  & Huygens' Quadrupole ESA  & 1.662 & 0.98 & 4.65   & 7.61  &  88.1  & 7.06 & Low\\

                  This work  & Huygens' Antenna  & 4.5 & 2.3578  & 10.1 & 8.37& 97 & 8.27 & Low\\

                         \hline
                    \hline
                \end{tabular}
                \label{tab1:comp}
            \end{center}

            \footnotesize{$^a$ The Harrington limit is used as the maximum gain; for an End-Fire array, another limit is based on the number of elements.\\
            $^b$ The use of extra-impedance matching networks and attenuators/amplifiers is assumed to constitute high complexity; only phase shifters/multi-layer designs form moderate complexity, while simple fabrication and no extra circuits form low complexity.}
            
 \end{table*}

For the TMM6-based design, the frequency was shifted downward due to the high permittivity, decreasing the overall electrical size. The achieved directivity is approximately 7.2 dBi and the maximum directivity with this antenna, corresponding to $ka$ = 1.85 according to Harrington (\ref{equ1}), is 8.52 dBi, introducing a gap of 1.32 dBi between the achieved and Harrington limits. For the design based on TMM3, the maximum directivity in the forward mode is 10.12 dBi, corresponding to $ka$ = 2.36. The achieved value is 8.37 dBi, resulting in a difference of approximately 1.74 dB. For the TMM3 backward mode, the maximum directivity is 9.07 dBi for $ ka = 2$, and the realized directivity is 7.4 dBi, resulting in a difference of approximately 1.67 dB. From this, it can be safely concluded that directivity approaches the Harrington limit as $ka$ approaches 1. For $ka <$ 1, the updated limits should be used, as discussed in the introduction. For $ka$ $>>$ 1, the directivity approaches that of the uniformly illuminated antenna, as given by (\ref{equ2}) and also discussed by Harrington \cite{harrington1958gain}.

It should be noted that the ESAs are not compared except in \cite{zhang2024superdirective} because the Harrington limit does not apply to these. For Huygens' ESAs, the maximum directivity limit is 4.8 dBi, which can only be surpassed by the excitation of multipoles \cite{zhang2024superdirective, ziolkowski2023superdirective}. In all these approaches, the gain cannot be scaled to a gain greater than 7.06 dBi for the Huygens' quadrupole antenna \cite{zhang2024superdirective}, except when bulky structures or large arrays are introduced. Here, using a low dielectric constant substrate such as RT/Duroid 5880, we demonstrated a gain of 9.33 dBi, which could be even higher without using any substrate and can easily reach beyond 10 dBi.

\section{Conclusion}
This paper demonstrates a superdirective antenna for secure and ultra-low power wireless communication using crossed electric and magnetic dipoles, forming a Huygens antenna. A magnetic dipole based on a half-wavelength slot is placed near a printed dipole that is twice the size of the slot. The induced magnetic dipole interacts with the electric dipole mode of the printed dipole; destructive interference occurs in the backward direction, and constructive interference occurs in the forward direction, resulting in a highly directive pattern. The proposed Huygens' antenna is designed to operate at 4.5 GHz, achieving a directivity of 8.37 dBi with a front-to-back ratio of 12 dB. The radiation efficiency of the antenna is 97\%, corresponding to a gain of 8.27 dBi. The high realized gain achieved from a compact planar structure with a high front-to-back ratio makes it suitable for several high-gain applications.
\bibliographystyle{IEEEtran}
\bibliography{ref}

\newpage

 




\vfill

\end{document}